\documentclass[12pt]{article}
\makeatletter
\def\section{\@startsection {section}{1}{\z@}{-3.5ex plus -1ex minus
 -.2ex}{2.3ex plus .2ex}{\large\bf}}
\def\subsection{\@startsection{subsection}{2}{\z@}{-3.25ex plus -1ex 
minus -.2ex}{1.5ex plus .2ex}{\normalsize\bf}}
\newcommand{\ky}{\kappa_{ \rm orb}}

\newcommand{\ket}[1]{|{#1}\rangle}
\newcommand{\braket}[2]{\langle{#1}|{#2}\rangle}
\newcommand{\be}{\begin{equation}}
\newcommand{\ee}{\end{equation}}
\begin{document}
\begin{titlepage}
\rightline{DFTT 37/2000}
\rightline{NORDITA-2000/95 HE}
\vskip 1.8cm
\centerline{\Large \bf FRACTIONAL D-BRANES AND}
\vskip 0.4cm
\centerline{\Large \bf  THEIR GAUGE DUALS 
\footnote{Work partially supported by the European Commission RTN
programme HPRN-CT-2000-00131 and by MURST.}}
\vskip 1.4cm 
\centerline{\bf M. Bertolini $^a$, P. Di Vecchia $^a$,
M. Frau $^b$, A. Lerda $^{c,b}$, 
R. Marotta $^a$, I. Pesando$^b$}
\vskip .8cm \centerline{\sl $^a$ NORDITA, Blegdamsvej 17, DK-2100
Copenhagen \O, Denmark}
\vskip .4cm \centerline{\sl $^b$ Dipartimento di  Fisica Teorica,
Universit\`a di Torino} \centerline{\sl and I.N.F.N., Sezione di
Torino,  Via P. Giuria 1, I-10125 Torino, Italy}
\vskip .4cm \centerline{\sl $^c$ Dipartimento di Scienze e Tecnologie
Avanzate} \centerline{\sl Universit\`a del Piemonte Orientale, I-15100
Alessandria, Italy}
\vskip 1.8cm
\begin{abstract}
We study the classical geometry associated to fractional D3-branes 
of type IIB string theory on ${\rm I\!R}_4/{\bf Z}_2$ which provide 
the gravitational dual for ${\cal{N}}=2$ super 
Yang-Mills theory in four dimensions.
As one can expect from the lack of conformal invariance on the gauge 
theory side, the gravitational background displays a repulson-like
singularity. It turns out however, that such singularity
can be excised by an enhan{\c{c}}on mechanism.
The complete knowledge of the classical supergravity solution
allows us to identify the coupling constant of the dual gauge theory 
in terms of the string parameters and to find a logarithmic 
running that is governed precisely by 
the $\beta$ function of the ${\cal N}=2$ super Yang-Mills theory.

\end{abstract}
\end{titlepage}

\renewcommand{\thefootnote}{\arabic{footnote}}
\setcounter{footnote}{0} \setcounter{page}{1} 

\tableofcontents  
\vskip 1cm
\section{Introduction}
After the seminal paper of 't Hooft on the large $N$ expansion~\cite{THOOFT},
many attempts to obtain a string theory out of QCD have been tried.
Recently, a remarkable progress in this direction has been achieved with the 
Maldacena conjecture \cite{MALDACENA} which states  
that ${\cal{N}}=4$ super Yang-Mills theory in four dimensions
is equivalent to type IIB string theory compactified on $AdS_5 \times S_5$.
This provides the first concrete example of how a string theory can be
extracted from a gauge theory. 
On the other hand, however, it was 
expected that a string model should emerge from a gauge field theory 
because of confinement, while the ${\cal{N}} =4$ super Yang-Mills 
is in the Coulomb phase.
Thus, a lot of effort has been recently devoted to extend the Maldacena 
conjecture and find new 
correspondences between string theories and
non-conformal and less supersymmetric gauge theories.
These attempts include the study of the renormalization group flow under a 
relevant perturbation in ${\cal{N}}=4$ super 
Yang-Mills~\cite{MILANO,ZAMORA,FREEDMAN}, the study of fractional branes on 
conifold singularities~\cite{KLEBA1,KLEBA2,KLEBA3,TATAR}, 
the study of the so-called ${\cal{N}}=1^*$ theory~\cite{POLCHISTRA},
and the search for the geometry of the stable non-BPS
D-branes~\cite{BERTOLINI}.

The common feature found in all these different examples is that
the classical geometric backgrounds have naked singularities
of repulson type. 
In some of these cases however, an interesting
phenomenon was discovered~\cite{JPP}:
a massive probe moving in these backgrounds becomes tensionless
before reaching the singularity. 
The geometric locus where this occurs is called enhan{\c{c}}on.
When this happens the supergravity approximation is not 
valid beyond the enhan{\c{c}}on and one is forced to consider 
stringy effects which should change the description and eventually excise
the singularity.
This phenomenon has been analyzed also in Refs.~\cite{BPP,PETRINI} 
for different configurations and in Ref. \cite{anto} for fractional D-branes
on K3 orbifolds. 
In Ref.~\cite{KLEBA3} however, it has been shown that
the repulson singularity of fractional branes on conifolds
can be removed already at the supergravity level
by suitably deforming the conifold, thus
obtaining a consistent gravitational dual that explains many features of
the gauge theory.
More recently, in Ref.~\cite{TSEYTLIN1} it has been
shown that instead the resolution of the conifold singularity 
is not sufficient to regularize the gravitational background and also
in this case an enhan{\c{c}}on mechanism seems to be necessary.
  
In this paper we study the classical geometry generated
by fractional D3-branes of type IIB string theory on the orbifold
${\rm I\!R}^{1,5}\,\times\,{\rm I\!R}^{4}/{\bf Z_2}$.
These are BPS configurations that are constrained to be at the orbifold 
fixed hyperplane and preserve eight supercharges.
The dual gauge theory corresponding to a stack of $M$ such fractional 
D3-branes is pure ${\cal N}=2$ super Yang-Mills theory in four dimensions 
with gauge group $U(M)$.
This is known to be non conformal and thus
it is interesting to check whether the dual classical
geometry displays a singular behavior.  Some features of this
solution indicating that this is indeed  the case were
already found in  Refs.~\cite{KLEBA1,GRANA}. In this paper we 
give the complete solution, with all physical quantities
expressed in terms of the string parameters $\alpha '$ and $g_s$, and
analyze its properties in some detail.

The first step for finding the exact solution is done using the 
boundary state formalism
that allows to determine
which supergravity fields are coupled to the brane and also their 
asymptotic behavior at large distance.
Using this formalism it is possible to infer the complete world-volume 
action for the fractional D3-branes and thus to obtain the non homogeneous 
equations that the supergravity fields must satisfy.
These equations can be explicitly solved and one can check 
that they describe configurations which satisfy the no-force condition 
as implied by the BPS bound.
As expected the classical solution exhibit a naked
singularity, which is in fact a repulson. Following the analysis done
in Refs.~\cite{JPP,BPP,PETRINI}, we see that the enhan\c{c}on mechanism
works also in our case. Hence the region of validity of the supergravity 
approximation  does not include the singularity.  
The properties of our solution  suggest a physical
picture  where the enhan\c{c}on geometry is  that of a ring like in 
Ref.~\cite{BPP,PETRINI}, instead of an  hypersphere as in
Ref.~\cite{JPP}. 

Exploiting the detailed knowledge of the solution of the fractional 
D3-brane and its world-volume action, we are able to find the
metric of the moduli space of the dual gauge theory and determine
from it the Yang-Mills coupling constant $g_{\rm YM}$ in terms of the string 
parameters. We find that $g_{\rm YM}$ is 
logarithmically running with a $\beta$-function that exactly 
matches the one of
${\cal{N}}=2$ super Yang-Mills theory in four dimensions.

\vskip 1cm

\section{The geometry of fractional D3-branes}

The action for type IIB supergravity in ten dimensions  can be written
(in the Einstein frame) as
\footnote{Our conventions for curved indices and forms are the
following:
$\varepsilon^{0\dots9}=+1$, signature $(-,+^9)$, $\omega_{(n)}={1\over n!}
\,\omega_{\mu_1 \dots \mu_n} dx^{\mu_1}\wedge\dots\wedge  dx^{\mu_n}$, and
$*\omega_{(n)}={\sqrt{-\det G}\over n!\,(10-n)!}\,
\varepsilon_{\nu_1\dots\nu_{10-n}\mu_1 \dots \mu_n} \,\omega^{\mu_1 \dots
\mu_n}  dx^{\nu_1}\wedge\dots\wedge dx^{\nu_{10-n}}.$}
\[
S_{\rm IIB} = \frac{1}{2 \kappa_{10}^2} \Bigg\{ \int d^{10} x~
 \sqrt{-\det G}~ R - \frac{1}{2} \int \Big[ d \phi \wedge {}^* d \phi
 \,+\, {\rm e}^{- \phi} H_{(3)}  \wedge {}^* H_{(3)}\,+\, {\rm e}^{2
 \phi}\, F_{(1)} \wedge {}^* F_{(1)}
\]
\begin{equation}
 + \,\,{\rm e}^{\phi} \,{\widetilde{F}}_{(3)} \wedge {}^*
 {\widetilde{F}}_{(3)} \,+\, \frac{1}{2}\, {\widetilde{F}}_{(5)}
 \wedge {}^* {\widetilde{F}}_{(5)}  \, +\,  C_{(4)} \wedge H_{(3)}
 \wedge F_{(3)} \Big] \Bigg\}
\label{tendim3}
\end{equation}
where \be H_{(3)} = d B_{(2)}~~~,~~~F_{(1)}=d C_{(0)}~~~,~~~ F_{(3)} =
d C_{(2)}~~~,~~~F_{(5)} = d C_{(4)}
\label{form2}
\ee are respectively the field strengths corresponding to the NS-NS
2-form potential, and to the 0-form, the  2-form and the 4-form
potentials of the R-R sector, and \be {\widetilde{F}}_{(3)} = F_{(3)}
- C_{(0)} \wedge H_{(3)}~~~~ ,~~~~{\widetilde{F}}_{(5)} = F_{(5)} -
C_{(2)} \wedge H_{(3)}~~.
\label{form3}
\ee 
Moreover,   $\kappa_{10}=8\,\pi^{7/2}\,g_s\,\alpha'^2$ where $g_s$
is the string coupling  constant, and the self-duality constraint
${}^* {\widetilde{F}}_{(5)}={\widetilde{F}}_{(5)}$  has to be
implemented on shell.  The D$p$-branes with $p$ odd are solutions of
the classical field equations that follow from the action
(\ref{tendim3}), which are charged under the R-R $(p+1)$-form
potentials and preserve sixteen supercharges. The D$3$-brane solution,
in which only the metric and the $4$-form potential  $C_{(4)}$ are
turned on, is particularly important because of the AdS/CFT
correspondence \cite{MALDACENA}.

Let us now consider type IIB supergravity on the orbifold  \be {\rm
I\!R}^{1,5}\,\times\,{\rm I\!R}^{4}/{\bf Z_2}
\label{orbifold}
\ee where ${\bf Z_2}$ is the reflection parity that changes the sign
to the four coordinates of ${\rm I\!R}^{4}$, which we take to be
$x^6$, $x^7$, $x^8$ and $x^9$. This is to be understood as the singular
limit of the corresponding ALE manifold.  The bulk action for this
theory is still given by eq.(\ref{tendim3}), but with $\kappa_{10}$
replaced by $\ky =\sqrt{2}\,\kappa_{10}=
(2\pi)^{7/2}\,g_s\,\alpha'^2$. In this case, besides the usual
D$p$-branes (bulk branes) which can freely move  in the orbifolded
directions, there are also {\it fractional}  D$p$-branes
\cite{FRACTIONAL} which are instead constrained to stay at the
orbifold fixed hyperplane $x^6=x^7= x^8=x^9=0$. These fractional
branes are the most elementary configurations of the theory, preserve
eight supercharges and can be viewed as D$(p+2)$-branes wrapped on the
(supersymmetric) vanishing 2-cycle of the orbifold.

In this paper we will consider in detail the fractional D$3$-brane.
From the supergravity point of view, this is a configuration  in
which the dilaton $\phi$ and the axion $C_{(0)}$ are constant, while
the metric, the 4-form $C_{(4)}$  and the two 2-forms $B_{(2)}$ and
$C_{(2)}$ are non-trivial.  More precisely, the latter fields, whose
presence is a distinctive feature of the fractional branes, are  \be
C_{(2)} = c~ \omega_2~~~~~,~~~~B_{(2)} = b~ \omega_2
\label{wra98}
\ee where $\omega_2$ is the $2$-form dual to the vanishing 2-cycle  of
the orbifold, and $c$ and $b$ are scalar fields living in  ${\rm
I\!R}^{1,5}$.

The fact that these are the non-trivial fields for a fractional
D$3$-brane  has a natural interpretation from a string  theory point
of view. In fact, let us consider the vacuum energy $Z$ between two
fractional D-branes which is given by the one-loop open string
amplitude
\begin{equation}
Z = \int_0^\infty \frac{d s}{ s} ~{\rm Tr}_{\rm NS-R} \left[\left(
\frac{1 + (-1)^F}{2}  \right) \left(\frac{1+g}{2} \right)  {\rm e}^{-
2 \pi s (L_0-a)} \right]
\label{part45}
\end{equation}
where $(-1)^F$ is the GSO parity, $g$ is the orbifold ${\bf Z}_2$
parity,  and the intercept is $a=1/2$ in the NS sector and $a=0$ in
the R sector.  By making the modular transformation $s \to 1/s$, one
can translate the one-loop open string amplitude (\ref{part45}) into a
tree-level closed string exchange diagram and, after factorization,
one can obtain the boundary state $\ket{B}$ associated to the
fractional  brane \cite{gabstef,BILLO}  (for a review of the boundary
state formalism and its applications see, for example,
\cite{antonella}).  The boundary state represents the source for the
closed strings emitted by the brane and in this case  it has four
different components which correspond to the (usual) NS-NS and R-R
untwisted sectors and to the NS-NS and R-R twisted sectors.  By
saturating the boundary state $\ket{B}$ with the massless closed
string states of the various sectors,  one can determine which are the
fields that couple to the fractional brane.  In particular, following
the procedure found in \cite{bs} and  reviewed in \cite{antonella},
one can find that in the untwisted sectors the fractional D3-brane
emits only the graviton $h_{\mu\nu}$
\footnote{We recall that the graviton field and the metric are related
by $G_{\mu\nu}=\eta_{\mu\nu} + 2\ky\,h_{\mu\nu}$.} and the 4-form
potential $C_{(4)}$. The couplings of these fields with the boundary
state are explicitly given by~\cite{anto} 
\be \braket{B}{h} =
-\frac{T_3}{\sqrt{2}}\,\, h_{\alpha}^{\,\,\, \alpha} \,V_4~~~,~~~
\braket{ B}{ C_{(4)}} = \frac{T_3}{\sqrt{2}\,\ky} \,C_{0123}\, V_4
\label{untw9}
\ee     
where $T_p =\sqrt{\pi} \,(2\pi\sqrt{\alpha '})^{(3-p)}$  is
the normalization of the boundary state, which is related to the brane
tension in units of the gravitational coupling constant \cite{bs},
$V_4$ is the (infinite) world-volume  of the D$3$-brane, and the index
$\alpha$ labels the longitudinal directions. By doing this same analysis
in the twisted sectors, we find that the boundary state of the
fractional D$3$-brane emits a massless scalar $\widetilde b$ from the
NS-NS sector and a $4$-form potential $A_{(4)}$ from the R-R
sector. Of course these fields exist only at the orbifold
fixed hyperplane $x^6=x^7=x^8=x^9=0$,  and thus their dynamics
develops in the remaining six-dimensional space.  The couplings of
these fields with the boundary state turn out to be given
by~\cite{anto} 
\be \braket{B}{\widetilde b} = -\frac{T_3}{\sqrt{2}\,\ky}
\,\frac{1}{2 \pi^2\alpha'}\,\widetilde b \, V_4 ~~~,~~~ \braket{B}{A_{(4)}} =
\frac{T_3}{\sqrt{2}\,\ky}\,\frac{1}{2 \pi^2 \alpha '}\,A_{0123}\, V_4~~.
\label{twi86}
\ee 
The twisted fields $\widetilde b$ and $A_{(4)}$ are related to the
fields $b$ and $c$ of eq.(\ref{wra98}). In fact, the scalar
$\widetilde b$ represents the fluctuation part of $b$ around the
background value which is characteristic of the ${\bf Z}_2$ orbifold
\cite{ASPINWALL,BILLO} 
\be b =  \frac{1}{2}\,{(4\pi^2\alpha')} +
\widetilde b~~,
\label{beta}
\ee 
while the potential $A_{(4)}$ is dual (in the six dimensional
sense) to the scalar $c$. To write down this duality  relation in a
correct way,  let us observe that the field equation for $C_{(2)}$
that follows from the action (\ref{tendim3}) is 
\be 
d\,{}^*dC_{(2)} =
F_{(5)}\wedge H_{(3)} = d\left(C_{(4)}\wedge H_{(3)}\right)~~,
\label{c2}
\ee 
so that we can write 
\be {}^*dC_{(2)} - C_{(4)}\wedge H_{(3)} =
dA_{(6)}~~.
\label{a6}
\ee 
The 6-form $A_{(6)}$ is the dual (in the ten dimensional sense) to
the R-R 2-form $C_{(2)}$. Let us now write this relation in the case
of the wrapped brane, {\it i.e.} using eq.(\ref{wra98}) and taking
$A_{(6)}= A_{(4)}\wedge\omega_2$. Then, one can easily find that 
\be
dA_{(4)} = - \,{}^{*_6}dc -C_{(4)}\wedge d\,b
\label{a4}
\ee 
where the Hodge dual $*_6$ is taken in the six-dimensional space
where the twisted fields live.

From the explicit couplings (\ref{untw9}) and (\ref{twi86}), it is
possible  to infer the form of the world-volume action of a fractional
D3-brane, namely
\begin{eqnarray}
S_{\rm boundary} &=& -\frac{T_3}{\sqrt{2}\,\ky} \int d^4x\,\sqrt{-\det
G_{\alpha\beta}} \left(1+\frac{1}{2\pi^2\alpha'}\, {\widetilde
b}\right)
\label{actionbound} \\
&&+\frac{T_3}{\sqrt{2}\,\ky} \int
C_{(4)}\left(1+\frac{1}{2\pi^2\alpha'}\, {\widetilde b} \right)  +
\frac{T_3}{\sqrt{2}\,\ky}\,\frac{1}{2\pi^2\alpha'} \int A_{(4)}~~~.
\nonumber
\end{eqnarray}
Of course, the boundary state calculations only determine the linear
terms of the world-volume action, but the higher order terms can be
found for example by imposing reparametrization invariance on the
world-volume (first line of (\ref{actionbound})) or by considering the
WZ part of the action of a D5-brane  wrapped on the (vanishing)
2-cycle in the presence of a non-trivial  $B_{(2)}$ field (second line
of (\ref{actionbound})).  The structure of the boundary action $S_{\rm
boundary}$ is confirmed also by explicit calculations of closed string
scattering amplitudes on a  disk with appropriate boundary conditions
\cite{MERLATTI}.

As explained in Ref.~\cite{bs},  the boundary state formalism allows
also to compute the asymptotic  behavior at large distance of the
various fields  in the classical brane solution.  For example, in our
fractional D3-brane we find that the metric is \be ds^2 \simeq
\left(1-\frac{Q}{2\,r^4}\right)\,\eta_{\alpha\beta} \,dx^\alpha
dx^\beta + \left(1+\frac{Q}{2\,r^4}\right)\,\delta_{ij}
\,dx^idx^j\,+...
\label{met1}
\ee where $\alpha,\beta=0,...,3$; $i,j=4,...,9$;
$r=\sqrt{x^ix^j\delta_{ij}}~$ and 
\be
Q\equiv\frac{\ky\,T_3}{2\sqrt{2}\,\pi^3} = 4\pi\, g_s\,\alpha'^2~~,
\label{Q}
\ee 
while the untwisted 4-form potential is 
\be C_{(4)} \simeq
-\,\frac{Q}{r^4}~dx^0\wedge dx^1\wedge dx^2\wedge dx^3 +...~~.
\label{c41}
\ee  The asymptotic behavior of the twisted fields is instead given by
\begin{eqnarray}
{\widetilde b} &\simeq& K\,\log (\rho/\epsilon)+...~~,
\label{b1} \\
A_{(4)} &\simeq& K\,\log (\rho/\epsilon)~dx^0\wedge dx^1\wedge dx^2
\wedge dx^3 +...~~,
\label{a41}
\end{eqnarray}
where $\rho = \sqrt{(x^4)^2 + (x^5)^2}$; $\epsilon$ is a regulator and
\be K \equiv \frac{\ky\,T_3}{\sqrt{2}\,\pi}~\frac{1}{2\pi^2\alpha'}
=4\pi\,g_s\,\alpha'~~.
\label{k}
\ee  
It is interesting to observe that while the untwisted fields
depend on the radial coordinate $r$ of the entire six-dimensional
transverse space, the twisted fields which do not see the four
orbifolded directions,  depend only on the  radial coordinate $\rho$
of the remaining two-dimensional transverse space.  This particular
feature was
also found in  Ref.~\cite{eyras} in the case of the non-BPS D-branes
in non-compact  orbifolds, while the logarithmic asymptotic behavior
of the twisted fields was already pointed out in Refs.~\cite{KLEBA1,TAKA}.

In the following  we look for an exact solution of the field equations of
type IIB supergravity with the asymptotic behavior described above.
We start by writing the equations of motion for the dilaton $\phi$ and
the  axion $C_{(0)}$, which are  
\be 
d {}^* d \phi = {\rm e}^{2 \phi}\,
dC_{(0)} \wedge {}^* dC_{(0)} +  \frac{1}{2}\,{\rm e}^{\phi}\,
{\widetilde{F}}_{(3)} \wedge {}^*{\widetilde{F}}_{(3)}  -
\frac{1}{2}\, {\rm e}^{-\phi}\, H_{(3)} \wedge {}^* H_{(3)}~~,
\label{dileq3}
\ee and \be d \left( {\rm e}^{2 \phi}\,{}^* d C_{(0)} \right) = -
\,{\rm e}^{\phi} {\widetilde{F}}_{(3)}  \wedge {}^* H_{(3)}~~.
\label{axion9}
\ee   As we discussed above, we are interested in a solution in which
both the  dilaton and the axion are constant, and the two 2-form
potentials are as  in eq.(\ref{wra98}).  To obtain this solution, it
is  convenient to introduce the combination~\footnote{Note 
that $G_{(3)}$ is not the $Sl(2,{\rm I\!R})$
invariant  3-form that is usually used in the supergravity literature, but
differs from the latter simply by a multiplicative factor.} 
\be
G_{(3)} = F_{(3)} - \tau H_{(3)}~~~{\rm with}~~~ \tau = C_{(0)} + {\rm
i}\, {\rm e}^{-\phi}~~.
\label{G3}
\ee  
For constant dilaton and axion,  eqs. (\ref{dileq3}) and (\ref{axion9})
imply that \be G_3 \wedge {}^*G_3 =0
\label{dilax4}
\ee which, using eq.(\ref{wra98}), in turn implies that \be d \gamma
\wedge {}^{*_6} d \gamma \wedge \omega_2 \wedge \omega_2 =0
\label{cond56}
\ee where we have defined the complex scalar \be \gamma = c
-\,\tau\,{\widetilde b}
\label{gamma}
\ee and taken into account the anti-selfduality of $\omega_2$.  If $
d\gamma \wedge {}^{*_6} d \gamma $ has components  along $x^4$ and
$x^5$, {\it i.e.} along the transverse directions  orthogonal to the
orbifold, then in order to satisfy eq.(\ref{cond56}) we must require
that \be \partial_z \gamma ~\partial_{\bar{z}} \gamma =0~~~{\rm
where}~~~z = x^4 +  {\rm i} \,x^5
\label{par43}
\ee 
which clearly can be satisfied by taking $\gamma$  to be, for
instance, an analytic function of  $z$~\cite{KLEBA1}. If we do this,
the dilaton and the axion can be consistently taken to be constant,
and, without any loss  of generality, we set them to zero.  With this
choice, of course we have $\tau={\rm i}~$.

Let us now turn to the other field equations. To derive them,  it is
convenient to first insert the {\it Ansatz} (\ref{wra98}) into the
original action  (\ref{tendim3}) and then use the fact that  the
integral of the product of two forms $\omega_2$ over the
four-dimensional orbifolded space is a constant that we choose so 
that the various fields in the bulk action have the canonical
normalization, apart from the overall factor of $1/ (2 \ky^2)$.
Proceeding in this way, we obtain the following action
\begin{eqnarray}
S'_{IIB} &=& \frac{1}{2 \ky^2} \Bigg\{ \int d^{10}x\, \sqrt{- \det
G}\, R -  \frac{1}{4}\int {\widetilde{F}}_{(5)} \wedge {}^*
{\widetilde{F}}_{(5)}  \nonumber \\ &&~~~~~~~~~~~~~~-~ \frac{1}{2}
\int  \left[ \,d {\bar{\gamma}} \wedge {}^{*_6} d \gamma -\frac{\rm
i}{2} C_{(4)}  \wedge d \gamma \wedge d {\bar{\gamma}} \,\right]_6
\Bigg\}
\label{sred5}
\end{eqnarray}
where the subindex $6$ in the second line indicates that the integral
is over the six-dimensional space orthogonal to the orbifolded
directions. At this point we can write the field
equations that follow from the total action  $S=S'_{\rm IIB} + S_{\rm
boundary}~$.
The equation for the 4-form potential 
$C_{(4)}$ is~\footnote{Note that, as usual, only the linear part of boundary 
action gives a non-trivial contribution to the field equations.} 
\be d \,{}^*
{\widetilde{F}}_{(5)} + \frac{\rm i}{2}\,  d \gamma \wedge d
{\bar{\gamma}}  \wedge {\Omega_4} + \left(\frac{2 \ky T_3}{\sqrt{2}}
\right)
{ \Omega_2}\wedge { \Omega_4} = 0
\label{f5}
\ee 
where we have defined
\begin{eqnarray}
{\Omega_4}&=& \delta(x^6)\cdots\delta(x^9)\,\,dx^6 \wedge\cdots\wedge
dx^9~~,\nonumber \\ {\Omega_2}&=& \delta(x^4)\,\delta(x^5)\,\,dx^4\wedge
dx^5~~,
\label{forms}
\end{eqnarray}
while the equation for the complex scalar $\gamma$ is 
\be \Big( d
\,{}^{*_6} d \gamma + \,{\rm i} \,{\widetilde{F}}_{(5)} \wedge  d
\gamma \Big) \wedge \Omega_4 +\,{\rm i}\, \left(\frac{2 \ky
T_3}{\sqrt{2}} \right) \frac{1}{2 \pi^2 \alpha '}  dx^0 \wedge
\cdots\wedge dx^{3}\wedge \Omega_2\wedge \Omega_4 = 0~~.
\label{gam74}
\ee 
Finally, the field equations for the metric are \be
{\widetilde{R}}_{\mu\nu} \equiv R_{\mu\nu} - \frac{1}{4 \,\cdot 4!}
({\widetilde{F}_{(5)}})_{\mu\lambda_1\ldots\lambda_4}
({\widetilde{F}_{(5)}})_{\nu}^{~\,\lambda_1\ldots\lambda_4} =
L_{\mu\nu}
\label{met54}
\ee 
where  
\be  L_{\alpha\beta}=-\frac{L}{\sqrt{-\det
G}}\,G_{\alpha\beta} \,~~~,~~~L_{ij}=\frac{L}{\sqrt{-\det G}}\,G_{ij}~~,
\label{L}
\ee 
with 
\be L=\left(\frac{1}{8} \sqrt{-\det
G_6}~\partial\gamma\cdot\partial{\bar\gamma} +\frac{\ky
T_3}{2\sqrt{2}}\sqrt{-\det G_{\alpha\beta}}~\delta(x^4)\,\delta(x^5)
\right)\delta(x^6)\cdots\delta(x^9) ~~.
\label{LL}
\ee 
We remark that in writing
eq.(\ref{LL}) we have used the analyticity of $\gamma$, and have
denoted by $G_6$ the metric in the six-dimensional space orthogonal to
the orbifold.

We now solve eqs.(\ref{f5}), (\ref{gam74}) and (\ref{met54})
by using a 3-brane-like {\rm Ansatz} for the untwisted fields, namely
\begin{eqnarray}
ds^2 &=& H^{-1/2}\, \eta_{\alpha\beta}\,d x^\alpha dx^\beta
+ H^{1/2} \,\delta_{ij} \,dx^i dx^j~~,
\label{met48} \\
{\widetilde{F}}_{(5)} &=&  d
\left(H^{-1} \, dx^0 \wedge \dots \wedge dx^3 \right)+ {}^* d
\left(H^{-1} \, dx^0 \wedge \dots \wedge dx^3 \right)~~.
\label{f5ans}
\end{eqnarray}
Inserting these expressions into eq.(\ref{gam74}), we easily obtain
\be
\delta^{ab}\,\partial_a\partial_b\,\gamma + {\rm i}\,\frac{2\ky T_3}{\sqrt{2}}
\,\frac{1}{2\pi^2\alpha'}\,\delta(x^4)\,\delta(x^5) = 0
\label{eqgamma}
\ee
where $a,b=4,5$, whose analytic solution is
\be
\gamma = -\,{\rm i} \,K \,\log (z/\epsilon)
\label{gammafin}
\ee
where $K$ is defined in eq.(\ref{k}) and $z=x^4+{\rm i}\,x^5$.
Taking the real and imaginary parts of $\gamma$, we get the twisted scalars
\begin{eqnarray}
c&=& K\,\tan^{-1}({x^5}/{x^4})
~~,\label{cfin}\\
{\widetilde b} &=& K\,\log(\rho/\epsilon)
~~.\label{bfin}
\end{eqnarray}
It is interesting to see that the asymptotic behavior
of ${\widetilde b}$ given in eq.(\ref{b1}) coincides with the 
complete solution (\ref{bfin}). Furthermore, using the duality relation
(\ref{a4}) and the {\it Ansatz} (\ref{met48})-(\ref{f5ans}), 
we can obtain the classical profile of the twisted
R-R potential $A_{(4)}$ appearing in the boundary action of the
fractional D3-brane. In fact, we have 
\be
A_{(4)} = K\,\log(\rho/\epsilon)\,dx^0\wedge dx^1\wedge dx^2
\wedge dx^3~~.
\label{a4fin}
\ee
Again the asymptotic form (\ref{a41}) 
obtained from the boundary state coincides with the full solution
(\ref{a4fin}).

Let us now find the equation that determines the warp factor $H$.
Inserting the {\it Ansatz} (\ref{met48})-(\ref{f5ans}) into
eq.(\ref{f5}), we get 
\be 
\delta^{ij}\,\partial_i
\partial_j H + |\partial_z \gamma|^2 \,\delta(x^6)\ldots\delta(x^9)
+\frac{2\ky T_3}{\sqrt{2}}\,\delta(x^4)\ldots \delta(x^9)=0 ~~.
\label{eq78}
\ee 
The last contribution is the standard source term that is present also for the
usual bulk D3-branes, while the second contribution is a peculiar
feature of the fractional D3-branes and represents the fact that,
in this case, the non-trivial flux of $G_{(3)}$ is a source for the
untwisted fields \cite{GRANA,CVETIC}. 
The final consistency check is to show that eq.
(\ref{eq78}) follows also from the Einstein equation (\ref{met54}). 
This is indeed what happens; in fact, using our {\it Ansatz},
it is possible to show that the left-hand side of eq.(\ref{met54})
becomes
\be
{\widetilde{R}}_{\alpha \beta} = \frac{\delta^{ij}\partial_i
\partial_j H }{4\, H^2}~\eta_{\alpha\beta}~~~,~~~{\widetilde{R}}_{ij} = -\,
\frac{\delta^{l k}\partial_l \partial_k H}{4\, H}~\delta_{ij}~~,
\label{tildeR}
\ee 
while the right-hand side becomes
\begin{eqnarray}
L_{\alpha \beta} &=& -\frac{1}{4\, H^2}\left(
|\partial_z \gamma|^2\,\delta(x^6)\ldots\delta(x^9)
+\frac{2\ky T_3}{\sqrt{2}}
\,\delta(x^4)\ldots \delta(x^9)\right)\,\eta_{\alpha\beta}~~,
\nonumber\\
L_{ij} &=& \frac{1}{4\,H}\left(
|\partial_z \gamma|^2\,\delta(x^6)\ldots\delta(x^9)
+\frac{2\ky T_3}{\sqrt{2}}
\,\delta(x^4)\ldots \delta(x^9)\right)\,\delta_{ij}~~.
\label{l99}
\end{eqnarray}
Hence, also the Einstein equation (\ref{met54}) implies eq.(\ref{eq78}).
Using standard techniques, it is possible to find the explicit solution
of this equation, and $H$ reads
\be 
H = 1+ \frac{Q}{r^4}  + \frac{2 K^2}{r^4} 
\left[\log \Bigg(\frac{r^4}{\epsilon^2 (r^2-\rho^2)}\Bigg) 
- 1+ \frac{\rho^2 }{r^2-\rho^2} \right]~~.
\label{H74}
\ee 
This expression is clearly in agreement with the results (\ref{met1})
and (\ref{c41})
obtained from the boundary state. 

Having the explicit form of the solution, we can now analyze its
properties. First of all, we can check that it respects the
no-force condition, as it should be because of its BPS properties.
To see this, we can substitute our classical solution
into the boundary action (\ref{actionbound}) and find 
\begin{eqnarray}
S_{\rm boundary} &=&-\frac{T_3\,V_4}{\sqrt{2}\,\ky}
\left\{H^{-1}\left[1+\frac{K}{2\pi^2\alpha'}\,\log(\rho/\epsilon)\right]
\nonumber\right. \\
&&\left.-~(H^{-1}-1)  \left[1+\frac{K}{2\pi^2\alpha'}\,
\log(\rho/\epsilon)\right]
-\frac{K}{2\pi^2\alpha'}\,\log(\rho/\epsilon)\right\}
\nonumber \\
&=&  -\frac{T_3\,V_4}{\sqrt{2}\,\ky}~~.
\end{eqnarray}
The fact that all position dependent terms exactly 
cancel leaving a constant
result is a check on the no-force condition to all orders; therefore, one
can safely form a stack of $M$ fractional D3-branes by simply
piling them on top of each other. In this case, the solution
for such a configuration has still the same form as before,
but with $Q\to M\,Q$ and $K\to M\,K$.

On the other hand, a closer look at the behavior of the
function $H$ in eq.(\ref{H74}) shows that the metric of the fractional 
D3-branes has a naked singularity. As we discussed in the 
introduction, this fact is a feature
that is shared also by other configurations which are dual to non-conformal
gauge theories~\cite{JPP,KLEBA2,BPP,PETRINI,TSEYTLIN1,BERTOLINI,anto}, 
and indeed
possess a naked singularity at some $r=r_0$. 
Actually, the structure of such singularity is that of a 
{\it repulson} because in its vicinity
the gravitational force, which is related to the gradient 
of the temporal component of the metric tensor, becomes
repulsive. Thus, there exists a region of anti-gravity and a distance
$r=r_e>r_0$ where the gravitational force vanishes.
A study of the shape of
$G_{00}$ indicates that the singularity is not a point in the
transverse six dimensional space but rather a two-dimensional surface.
In fact, the repulson is located near $x^6=x^7=x^8=x^9=0$ and extends along the
non-orbifolded transverse directions $x^4$ and $x^5$. Note
that the breaking of the spherical symmetry in the six-dimensional
transverse space is not surprising since the
starting vacuum geometry, eq.(\ref{orbifold}), already breaks it.
Moreover, a simple numerical analysis reveals that the singularity 
does not cover the full $x^4,x^5$ plane; in fact the temporal
component of the metric tensor ceases to be singular at some value
$\rho_0$ and smoothly goes to zero for bigger values of $\rho$, signaling
the possible appearance of an horizon. Clearly, a more detailed analytical
study of the classical geometry produced by a fractional D3-brane
and of its singularity is needed. Here we have just mentioned 
the most relevant features which are useful and sufficient for the discussion
of the following section.

\vskip 1cm
\section{The enhan\c{c}on and the dual gauge theory}

The discussion of the previous section and the presence in our solution
of a geometrical locus where the gravitational force vanishes, clearly
indicates the possibility that an {enhan\c{c}on} mechanism may take place
\cite{JPP}~\footnote{Notice that in \cite{JPP}, 
the {enhan\c{c}on} locus coincides with the
locus where the gravitational force vanishes, but
this may not be necessarily the case for more general configurations. The
"physical" enhan\c{c}on occurs, by definition, where the
probe brane becomes tensionless.}. This would excise the repulson from
the metric, thus
yielding a singularity-free solution.  

In order to see whether 
this really happens, we apply the same methods of Ref.~\cite{JPP}
and study the (slow) motion of a probe brane 
in the geometry generated by $M$ fractional D3-branes (for 
a review on this approach see, for example, \cite{JOHNSON}). 
This can be done by inserting the classical solution 
into the world-volume action (\ref{actionbound}) 
and expanding it in powers of the
velocity of the probe D3-brane in the two transverse directions $x^4$
and $x^5$. Doing this, we find that 
the position dependent terms cancel exactly because of supersymmetry,
as we have already seen in the previous section, 
while the terms quadratic in the 
velocity of the probe survive and allow to define a non-trivial two-dimensional
metric on moduli space. More precisely, from the DBI part of
boundary action, we get
\be
 \frac{1}{2}\,\frac{T_3}{\sqrt{2}{\ky}} \int d^4 x \,
\left( \frac{\partial x^a}{\partial x^0}\right)^2  
\left( 1 + \frac{\widetilde{b}}{2 \pi^2\alpha'} \right)
\label{probe34}
\ee 
where the index $a$ takes values $4$ and $5$.
We now express the various constants of (\ref{probe34})
in terms of the string parameters,
use eq.(\ref{bfin}) and identify
the coordinates $x^a = 2 \pi \alpha' \Phi^a$ with the Higgs fields, so that
the kinetic term for the scalars can be rewritten as follows 
\be 
- \frac{1}{2} \int d^4 x ~\partial_{\mu} \Phi^a
\partial^{\mu} \Phi^b \,g_{ab}
\label{kinte4}
\ee 
where the metric in moduli space is 
\be 
g_{ab} = \frac{1}{8 \pi g_s}\,\left(1 + \frac{M\,K}{2\pi^2\alpha'}  \log
(\rho/\epsilon) \right) \delta_{ab}~~.
\label{metmod}
\ee
It is easy to see that this metric vanishes when $\rho$ reaches
the following value
\be 
\frac{\rho_e}{\epsilon} = {\rm e}^{- \pi/(2 M g_s) }~~.
\label{en89}
\ee
This means that at $\rho =\rho_e$ the probe fractional brane becomes
tensionless; thus for $\rho < \rho_e$ 
the supergravity solution looses its meaning
because there the probe gets a negative tension. This fact can be
interpreted also as the signal that new degrees of freedom are becoming
massless below $\rho_e$ and that they  have to be suitably taken into
account with a fully stringy description. 
A phenomenon similar to the one 
originally discussed in \cite{JPP} is at work here: the true
microscopic configuration is not given by a stack of coincident branes
but rather by an hypersurface (defined by eq.(\ref{en89}))
on which the branes are smeared. 
However, differently  from
\cite{JPP}, our enhan{\c{c}}on is not an (hyper)sphere in the transverse
space, but rather a ring depending on the two coordinates, $x^4$ and $x^5$
through $\rho$.  
Another important point to notice is 
that at the enhan{\c{c}}on, the fluctuation part ${\widetilde b}$
of the twisted scalar field $b$ exactly 
cancels its background value which, in unit of the string length $2 \pi
\sqrt{\alpha'}$, is $1/2$ (see eqs.(\ref{beta}) and (\ref{bfin})). 

A more detailed characterization of the
structure of the regularized classical solution deserves
further study. Nevertheless, what we have found here is already
enough to get  interesting information about the dual field theory.
We recall that in the case of $M$ 
fractional D3-branes, the world-volume gauge theory is pure 
${\cal N}=2$ super Yang-Mills in four dimensions
with gauge group $U(M)$. This can be simply understood
by analyzing the massless spectrum of the open strings
attached to the fractional D3-branes. Notice that no
hypermultiplets are present since the corresponding
moduli would be related to displacements of the branes from the
orbifold fixed point, which are not possible for fractional
branes~\footnote{This is to be 
contrasted with the case of $N$ bulk branes where 
the gauge group is $U(N)\times U(N)$ and one expects
also hypermultiplets to be present~\cite{DOUGLMOORE}.}.

Remembering that the Higgs fields $\Phi^a$ are the two scalars of the
${\cal N}=2$ vector multiplet, from the action (\ref{kinte4}) we can
read that the Yang-Mills coupling constant is given by
\be
g_{\rm YM}^{2} (\mu) = 
({g_{\rm YM}^0})^{2}\,\left(1 
+  \frac{M \,({g_{\rm YM}^0})^{2} }{4 \pi^2}\,
\log \mu\right)^{-1}
\label{gaugecou98}
\ee
where
\be
({g_{\rm YM}^0})^{2}\equiv
g_{\rm YM}^{2} ( \mu=1 ) = 8 \pi g_s ~~~,~~~ \mu \equiv
\frac{\rho}{\epsilon}~~.
\ee
Eq.(\ref{gaugecou98}) defines the running of the YM coupling constant 
with the variation of the scale $\mu$. 
Notice that from this point of view, the enhan\c{c}on  
locus (\ref{en89}) defines the scale at 
which $g_{\rm YM}$ diverges. 
Remembering that $\rho\rightarrow \infty$ corresponds 
to the ultraviolet limit in the dual field theory, we see that 
the coupling constant (\ref{gaugecou98}) describes an asymptotically
free gauge theory! Finally, by computing the 
$\beta$-function, we find
\be
\beta \equiv \mu \frac{\partial}{\partial \mu} 
g_{\rm YM}(\mu) = - \frac{g_{\rm YM}^{3}(\mu)}{8 
\pi^2} \,M
\label{betafu}
\ee 
which is precisely the $\beta$-function of the pure 
${\cal{N}}=2$ super Yang-Mills theory (modulo instanton corrections).
It would be interesting to investigate the relation
between our results and those of Ref.~\cite{ANSAR}
where the ${\cal N}=2$ gauge theories
are obtained from M5-branes of 11-dimensional supergravity
wrapped on Riemann surfaces.

It is also worth pointing out that the R-R twisted scalar $c$
of eq.(\ref{cfin}) is directly related to the $\theta$-angle of
the YM theory. In fact, by introducing a gauge field $F$ in the
world-volume action of the probe D3-brane and expanding the WZ part
in powers of $F$, we can read from the
coefficient in front of the ${\rm Tr}(F\wedge F)$ term that
\be
c= 2\pi\,\alpha'\,g_s\,\theta_{YM}~~.
\label{theta}
\ee
As a consequence, the complex scalar ${\bar \gamma}=c+{\rm i}\,b$
of the supergravity solution can be nicely written as
\be
{\bar \gamma}=(2\pi\sqrt{\alpha'})^2\,g_s\,\tau
\ee
where $\tau$ is the standard combination of the
YM coupling constant and $\theta$-angle
\be
\tau = \frac{\theta_{YM}}{2\pi}+{\rm i}\,\frac{4\pi}{g_{YM}^2}~~.
\ee

We conclude by observing that it is straightforward to
extend our analysis to the case in which there are 
$N$ bulk D3-branes besides the $M$ fractional ones 
considered so far. The only change in the classical
solution corresponding to this configuration occurs in the
function $H$ of eq.(\ref{H74}) in which the parameter $Q$ 
must be replaced by $(2N+M)\,Q$, while $K$ must be replaced by
$M\,K$ as before. 
Furthermore, the enhan{\c{c}}on locus gets changed to
\be 
\frac{\rho_e}{\epsilon} = {\rm e}^{- \pi(1+2n)/(2 M g_s) }
\label{en899}
\ee
where $n=N/M$.

Our results show that is possible to obtain precise non-perturbative
information on a gauge theory using the dual classical geometry
provided by D-branes, even in cases different from those of the
AdS/CFT correspondence. 
This fact hints the possibility that the gauge/gravity duality 
has an even deeper meaning than expected. 


\vskip 1.5cm 
{\large {\bf Acknowledgments}}
\vskip 0.5cm
\noindent
We would like to thank M. Bill\'o, P. Fr\`e, L. Gallot, A. Liccardo, R. Musto, 
M. Petrini and R. Russo for very useful discussions. M.B. acknowledges
support from INFN.


\begin{thebibliography}{99}

\bibitem{THOOFT} G. 't Hooft, Nucl. Phys. {\bf B72} (1974) 461.

\bibitem{MALDACENA}J. Maldacena,  Adv. Theor. Math. Phys. {\bf 2} 
(1998) 231, {\tt hep-th/9711200}.

\bibitem{MILANO}
L. Girardello, M. Petrini, M. Porrati and A. Zaffaroni, JHEP 9812 
(1998) 022, {\tt hep-th/9810126}; 
JHEP 9905 (1999) 026, {\tt hep-th/9903026}; 
Nucl. Phys {\bf B569} (2000) 451, {\tt hep-th/9909047}.

\bibitem{ZAMORA} J. Distler and F. Zamora, Adv. Theor. Math. Phys. {\bf 2} 
(1998) 1405, {\tt hep-th/
9810206}.

\bibitem{FREEDMAN} D.Z. Freedman, S.S. Gubser, K. Pilch and N.P. Warner, 
{\em Renormalization Group Flows from Holography--Supersymmetry and 
a c-Theorem}, {\tt hep-th/9904017}; 
JHEP {\bf 0007} (2000) 038, {\tt hep-th/9906194}.

\bibitem{KLEBA1} I.R. Klebanov and N. Nekrasov, 
Nucl. Phys. {\bf B574} (2000) 263, {\tt hep-th/  
9911096}.

\bibitem{KLEBA2} I.R. Klebanov and A.A. Tseytlin,
Nucl. Phys. {\bf B578} (2000) 123, {\tt hep-th/
0002159}.

\bibitem{KLEBA3} I.R. Klebanov and M.J. Strassler, 
JHEP {\bf 0008} (2000) 052, {\tt hep-th/0007191}.

\bibitem{TATAR}K. Oh and R. Tatar, JHEP {\bf 0005} (2000) 030, 
{\tt hep-th/0003183}.

\bibitem{POLCHISTRA} J. Polchinski and M.J. Strassler, 
{\em The String Dual of a Confining Four-Dimensional Gauge Theory},
{\tt hep-th/0003136}.

\bibitem{BERTOLINI} M. Bertolini, P. Di Vecchia, M. Frau, A. Lerda, R. Marotta
and R. Russo, Nucl. Phys. {\bf B590} (2000) 471, {\tt hep-th/0007097}.

\bibitem{JPP} C.V. Johnson, A.W. Peet and J. Polchinski, 
Phys. Rev {\bf D61} (2000) 086001, {\tt hep-th/9911161}.

\bibitem{BPP} A. Buchel, A.W. Peet and J. Polchinski, 
{\em Gauge Dual and Noncommutative Extension of an N=2 Supergravity Solution},
{\tt hep-th/0008076}.

\bibitem{PETRINI} N. Evans, C.V. Johnson and M. Petrini, 
{\em The Enhan{\c{c}}on and N=2 Gauge Theory/Gravity RG Flows},
{\tt hep-th/0008081}.

\bibitem{anto}M. Frau, A. Liccardo, R. Musto, {\em The geometry of
fractional branes}, {\tt hep-th/
0012035}.

\bibitem{TSEYTLIN1} L.A. Pando Zayas and A.A. Tseytlin,
{\em 3-branes on resolved conifold}, {\tt hep-th/
0010088}.

\bibitem{GRANA} M. Grana and J. Polchinski, 
{\em Supersymmetric Three-Form Flux Perturbations on $AdS_5$},
{\tt hep-th/0009211}.

\bibitem{FRACTIONAL}M.R. Douglas, JHEP {\bf 9707} (1997) 004;
D. Diaconescu, M.R. Douglas and J. Gomis, JHEP {\bf 9802} (1998) 013,
{\tt hep-th/9712230}.

\bibitem{gabstef}
D. Diaconescu and J. Gomis, JHEP {\bf 0010} (2000) 001,
{\tt hep-th/9906242}; 
M.R. Gaberdiel and J.B.~Stefanski, Nucl. Phys. {\bf B578}
(2000) 58, {\tt hep-th/9910109}.

\bibitem{BILLO} M. Bill\`o, B. Craps and F. Roose,
{\em Orbifold boundary states from Cardy's condition}, 
{\tt hep-th/0011060}.

\bibitem{antonella}
P.~Di Vecchia and A.~Liccardo, {\em {D}-branes in string theories I\/},
{\tt hep-th/9912161}; 
{\em {D}-branes in string theories II\/}, {\tt hep-th/9912275}.

\bibitem{bs}
P.~Di Vecchia, M.~Frau, I.~Pesando, S.~Sciuto, A.~Lerda and R.~Russo, Nucl.
Phys. {\bf B507} (1997) 259, {\tt hep-th/9707068}; P. Di Vecchia,
M. Frau, A. Lerda and A. Liccardo, Nucl. Phys. {\bf B565} (2000) 397,
{\tt hep-th/9906214}.

\bibitem{ASPINWALL}P. Aspinwall, 
{\em K3 surfaces and string duality}, {\tt hep-th/9611137}.

\bibitem{MERLATTI} P. Merlatti and G. Sabella,
{\em World Volume Action for Fractional Branes}, {\tt hep-th/0012193}.

\bibitem{eyras}
E.~Eyras and S.~Panda, Nucl. Phys. {\bf B584} (2000) 251,
{\tt hep-th/0003033}.

\bibitem{TAKA}T. Takayanagi, JHEP {\bf 0002} (2000) 040, {\tt hep-th/9912157}.

\bibitem{CVETIC}M. Cvetic, H. Lu and C.N. Pope, {\em Brane resolution
through trasgression}, {\tt hep-th/0011023}.

\bibitem{JOHNSON}C.V. Johnson, {\em D-brane primer}, {\tt hep-th/0007170}.

\bibitem{DOUGLMOORE}M.R. Douglas and G. Moore, 
{\em Branes, Quivers, and ALE Instantons}, {\tt hep-th/
9603167}.

\bibitem{ANSAR} 
A. Fayyazuddin and D.J. Smith, JHEP {\bf 0010} (2000) 023, 
{\tt hep-th/0006060}; B. Brinne, A. Fayyazuddin, S. Mukhopadhyay and
D.J. Smith, {\em Supergravity M5-branes wrapped on Riemann
surfaces and their QFT duals}, {hep-th/0009047}.

\end{thebibliography}
\end{document}